
\message{JNL.TEX version 0.95 as of 10/30/86.}

\font\twelverm=cmr12                    \font\twelvei=cmmi12
\font\twelvesy=cmsy10 scaled 1200       \font\twelveex=cmex10 scaled 1200
\font\twelvebf=cmbx12                   \font\twelvesl=cmsl12
\font\twelvett=cmtt12                   \font\twelveit=cmti12
\font\twelvesc=cmcsc10 scaled 1200
\skewchar\twelvei='177                  \skewchar\twelvesy='60


\def\twelvepoint{\normalbaselineskip=12.4pt plus 0.1pt minus 0.1pt
  \abovedisplayskip 12.4pt plus 3pt minus 9pt
  \belowdisplayskip 12.4pt plus 3pt minus 9pt
  \abovedisplayshortskip 0pt plus 3pt
  \belowdisplayshortskip 7.2pt plus 3pt minus 4pt
  \smallskipamount=3.6pt plus1.2pt minus1.2pt
  \medskipamount=7.2pt plus2.4pt minus2.4pt
  \bigskipamount=14.4pt plus4.8pt minus4.8pt
  \def\rm{\fam0\twelverm}          \def\it{\fam\itfam\twelveit}%
  \def\sl{\fam\slfam\twelvesl}     \def\bf{\fam\bffam\twelvebf}%
  \def\mit{\fam 1}                 \def\cal{\fam 2}%
  \def\sc{\twelvesc}               \def\tt{\twelvett}
  \textfont0=\twelverm   \scriptfont0=\tenrm   \scriptscriptfont0=\sevenrm
  \textfont1=\twelvei    \scriptfont1=\teni    \scriptscriptfont1=\seveni
  \textfont2=\twelvesy   \scriptfont2=\tensy   \scriptscriptfont2=\sevensy
  \textfont3=\twelveex   \scriptfont3=\twelveex  \scriptscriptfont3=\twelveex
  \textfont\itfam=\twelveit
  \textfont\slfam=\twelvesl
  \textfont\bffam=\twelvebf \scriptfont\bffam=\tenbf
  \scriptscriptfont\bffam=\sevenbf
  \normalbaselines\rm}


\def\tenpoint{\normalbaselineskip=12pt plus 0.1pt minus 0.1pt
  \abovedisplayskip 12pt plus 3pt minus 9pt
  \belowdisplayskip 12pt plus 3pt minus 9pt
  \abovedisplayshortskip 0pt plus 3pt
  \belowdisplayshortskip 7pt plus 3pt minus 4pt
  \smallskipamount=3pt plus1pt minus1pt
  \medskipamount=6pt plus2pt minus2pt
  \bigskipamount=12pt plus4pt minus4pt
  \def\rm{\fam0\tenrm}          \def\it{\fam\itfam\tenit}%
  \def\sl{\fam\slfam\tensl}     \def\bf{\fam\bffam\tenbf}%
  \def\smc{\tensmc}             \def\mit{\fam 1}%
  \def\cal{\fam 2}%
  \textfont0=\tenrm   \scriptfont0=\sevenrm   \scriptscriptfont0=\fiverm
  \textfont1=\teni    \scriptfont1=\seveni    \scriptscriptfont1=\fivei
  \textfont2=\tensy   \scriptfont2=\sevensy   \scriptscriptfont2=\fivesy
  \textfont3=\tenex   \scriptfont3=\tenex     \scriptscriptfont3=\tenex
  \textfont\itfam=\tenit
  \textfont\slfam=\tensl
  \textfont\bffam=\tenbf \scriptfont\bffam=\sevenbf
  \scriptscriptfont\bffam=\fivebf
  \normalbaselines\rm}


\def\beginlinemode{\endmode
  \begingroup\parskip=0pt \obeylines\def\\{\par}\def\endmode{\par\endgroup}}
\def\beginparmode{\endmode
  \begingroup \def\endmode{\par\endgroup}}
\let\endmode=\par
{\obeylines\gdef\
{}}
\def\percentspace{\baselineskip=\normalbaselineskip\divide\baselineskip by 100
  \multiply\baselineskip by}
\def\singlespace        {\percentspace100}

\def\oneandahalfspace   {\percentspace150}
\def\doublespace        {\percentspace200}

\newcount\firstpageno\firstpageno=2
\footline={\ifnum\pageno<\firstpageno{\hfil}\else{\hfil\twelverm\folio\hfil}\fi}
\def\toppageno{\global\footline={\hfil}\global\headline
  ={\ifnum\pageno<\firstpageno{\hfil}\else{\hfil\twelverm\folio\hfil}\fi}}
\let\rawfootnote=\footnote              
\def\footnote#1#2{{\rm\singlespace\parindent=0pt\parskip=0pt
  \rawfootnote{#1}{#2\hfill\vrule height 0pt depth 6pt width 0pt}}}

\def\raggedcenter{\leftskip=4em plus 12em \rightskip=\leftskip
  \parindent=0pt \parfillskip=0pt \spaceskip=.3333em \xspaceskip=.5em
  \pretolerance=9999 \tolerance=9999
  \hyphenpenalty=9999 \exhyphenpenalty=9999 }
\def\dateline{\rightline{\ifcase\month\or
  January\or February\or March\or April\or May\or June\or
  July\or August\or September\or October\or November\or December\fi
  \space\number\year}}
\def\received{\vskip 3pt plus 0.2fill
 \centerline{\sl (Received\space\ifcase\month\or
  January\or February\or March\or April\or May\or June\or
  July\or August\or September\or October\or November\or December\fi
  \qquad, \number\year)}}


\hsize=6.5truein
\hoffset=0.15truein
\vsize=8.9truein
\voffset=0.1truein
\def\landscape{ \hsize=8.9truein\hoffset=0.1truein
                \vsize=6.5truein\voffset=0.15truein
                \xdef\endit{\immediate\write16
 {>>>>To print this, use the command IMPRINT/LAND/DVI
\jobname.DVI<<<<.}\endit}}
\parskip=\medskipamount
\def\\{\cr}
\twelvepoint            
\doublespace            
\overfullrule=0pt       


\def\timestamp{\input timestamp \timestamp}     


\def\title                      
  {\null\vskip 3pt plus 0.2fill
   \beginlinemode \doublespace \raggedcenter \bf}

\def\author                     
  {\vskip 3pt plus 0.2fill \beginlinemode
   \singlespace \raggedcenter\sc}

\def\affil                      
  {\vskip 3pt plus 0.1fill \beginlinemode
   \oneandahalfspace \raggedcenter \sl}

\def\abstract                   
  {\vskip 3pt plus 0.3fill \beginparmode
   \oneandahalfspace ABSTRACT: }

\def\endtitlepage               
  {\endpage                     
   \body}

\def\body                       
  {\beginparmode}               

\def\head#1{                    
  \goodbreak\vskip 0.5truein    
  {\immediate\write16{#1}
   \raggedcenter \uppercase{#1}\par}
   \nobreak\vskip 0.25truein\nobreak}

\def\refto#1{$^{#1}$}           

\def\references                 
  {\head{References}            
   \beginparmode
   \frenchspacing \parindent=0pt \leftskip=1truecm
   \parskip=8pt plus 3pt \everypar{\hangindent=\parindent}}

\gdef\refis#1{\item{#1.\ }}                     


\def\endreferences{\body}

\def\figure#1. (#2) #3\par{     
  \goodbreak\midinsert
  \hskip-\hoffset\noindent\null\vskip #2\relax
  {\tenpoint\singlespace\leftskip=1in\rightskip=0.5in
  \noindent Figure #1. #3\par}\endinsert}

\def\figurefile#1. (#2) #3\par{ 
  \goodbreak\midinsert
  \hskip-\hoffset\noindent\special{insert #2}\null\vskip4.75in
  {\tenpoint\singlespace\leftskip=1in\rightskip=0.5in
  \noindent Figure #1. #3\par}\endinsert}

\def\figurecaptions             
  {\endpage
   \beginparmode
   \head{Figure Captions}
}

\def\endfigurecaptions{\body}

\def\endpage                    
  {\vfill\eject}

\def\endpaper                   
  {\endmode\vfill\supereject}

\def\endit
  {\endpaper\end}


\def\heading                            
  {\vskip 0.5truein plus 0.1truein      
   \beginparmode \def\\{\par} \parskip=0pt \singlespace \raggedcenter}

\def\subheading                         
  {\vskip 0.25truein plus 0.1truein     
   \beginlinemode \singlespace \parskip=0pt \def\\{\par}\raggedcenter}

\def\tag#1$${\eqno(#1)$$}

\def\align#1$${\eqalign{#1}$$}

\def\aligntag#1$${\gdef\tag##1\\{&(##1)\cr}\eqalignno{#1\\}$$
  \gdef\tag##1$${\eqno(##1)$$}}

\def\overset #1\to#2{{\mathop{#2}\limits^{#1}}}
\def\underset#1\to#2{{\let\next=#1\mathpalette\undersetpalette#2}}
\def\undersetpalette#1#2{\vtop{\baselineskip0pt
\ialign{$\mathsurround=0pt #1\hfil##\hfil$\crcr#2\crcr\next\crcr}}}


\def\ref#1{Ref.~#1}                     
\def\Ref#1{Ref.~#1}                     
\def\[#1]{[\cite{#1}]}
\def\cite#1{{#1}}
\let\eq=\Eq                
\def\(#1){(\call{#1})}
\def\call#1{{#1}}
\def\taghead#1{}
\def\frac#1#2{{#1 \over #2}}

\def\1{\frac1}\def\2{\frac2}\def\3{\frac3}\def\4{\frac4}\def\5{\frac5}
\def\6{\frac6}\def\7{\frac7}\def\8{\frac8}\def\9{\frac9}

\def\ie{{\it i.e.,\ }}

\def\etal{{\it et al.\ }}
\def\etc{{\it etc.\ }}

\def\sla{\raise.15ex\hbox{$/$}\kern-.57em}
\def\leaderfill{\leaders\hbox to 1em{\hss.\hss}\hfill}
\def\twiddle{\lower.9ex\rlap{$\kern-.1em\scriptstyle\sim$}}
\def\bigtwiddle{\lower1.ex\rlap{$\sim$}}
\def\gtwid{\mathrel{\raise.3ex\hbox{$>$\kern-.75em\lower1ex\hbox{$\sim$}}}}
\def\ltwid{\mathrel{\raise.3ex\hbox{$<$\kern-.75em\lower1ex\hbox{$\sim$}}}}
\def\square{\kern1pt\vbox{\hrule height 1.2pt\hbox{\vrule width 1.2pt\hskip 3pt
   \vbox{\vskip 6pt}\hskip 3pt\vrule width 0.6pt}\hrule height 0.6pt}\kern1pt}
\def\tdot#1{\mathord{\mathop{#1}\limits^{\kern2pt\ldots}}}

\def\beneathrel#1\under#2{\mathrel{\mathop{#2}\limits_{#1}}}
\def\pmb#1{\setbox0=\hbox{#1}%
  \kern-.025em\copy0\kern-\wd0
  \kern  .05em\copy0\kern-\wd0
  \kern-.025em\raise.0433em\box0 }

\def\beginitems{\par\begingroup\parskip=0pt\advance\leftskip by 24pt
  \def\i##1 {\item{##1}}\def\ii##1 {\itemitem{##1}}\medskip}

\catcode`@=11
\newcount\r@fcount \r@fcount=0
\newcount\r@fcurr
\immediate\newwrite\reffile
\newif\ifr@ffile\r@ffilefalse
\def\w@rnwrite#1{\ifr@ffile\immediate\write\reffile{#1}\fi\message{#1}}

\def\writer@f#1>>{}
\def\referencefile{
  \r@ffiletrue\immediate\openout\reffile=\jobname.ref%
  \def\writer@f##1>>{\ifr@ffile\immediate\write\reffile%
    {\noexpand\refis{##1} = \csname r@fnum##1\endcsname = %
     \expandafter\expandafter\expandafter\strip@t\expandafter%
     \meaning\csname r@ftext\csname r@fnum##1\endcsname\endcsname}\fi}%
  \def\strip@t##1>>{}}

\def\citeall#1{\xdef#1##1{#1{\noexpand\cite{##1}}}}
\def\cite#1{\each@rg\citer@nge{#1}}     

\def\each@rg#1#2{{\let\thecsname=#1\expandafter\first@rg#2,\end,}}
\def\first@rg#1,{\thecsname{#1}\apply@rg}       
\def\apply@rg#1,{\ifx\end#1\let\next=\relax
\else,\thecsname{#1}\let\next=\apply@rg\fi\next}

\def\citer@nge#1{\citedor@nge#1-\end-}  
\def\citer@ngeat#1\end-{#1}
\def\citedor@nge#1-#2-{\ifx\end#2\r@featspace#1 
  \else\citel@@p{#1}{#2}\citer@ngeat\fi}        
\def\citel@@p#1#2{\ifnum#1>#2{\errmessage{Reference range #1-#2\space is bad.}%
    \errhelp{If you cite a series of references by the notation M-N, then M and
    N must be integers, and N must be greater than or equal to M.}}\else%
 {\count0=#1\count1=#2\advance\count1
by1\relax\expandafter\r@fcite\the\count0,%
  \loop\advance\count0 by1\relax
    \ifnum\count0<\count1,\expandafter\r@fcite\the\count0,%
  \repeat}\fi}

\def\r@featspace#1#2 {\r@fcite#1#2,}    
\def\r@fcite#1,{\ifuncit@d{#1}
    \newr@f{#1}%
    \expandafter\gdef\csname r@ftext\number\r@fcount\endcsname%
                     {\message{Reference #1 to be supplied.}%
                      \writer@f#1>>#1 to be supplied.\par}%
 \fi%
 \csname r@fnum#1\endcsname}
\def\ifuncit@d#1{\expandafter\ifx\csname r@fnum#1\endcsname\relax}%
\def\newr@f#1{\global\advance\r@fcount by1%
    \expandafter\xdef\csname r@fnum#1\endcsname{\number\r@fcount}}

\let\r@fis=\refis                       
\def\refis#1#2#3\par{\ifuncit@d{#1}
   \newr@f{#1}%
   \w@rnwrite{Reference #1=\number\r@fcount\space is not cited up to now.}\fi%
  \expandafter\gdef\csname r@ftext\csname r@fnum#1\endcsname\endcsname%
  {\writer@f#1>>#2#3\par}}

\def\ignoreuncited{
   \def\refis##1##2##3\par{\ifuncit@d{##1}%
     \else\expandafter\gdef\csname r@ftext\csname
r@fnum##1\endcsname\endcsname%
     {\writer@f##1>>##2##3\par}\fi}}

\def\r@ferr{\endreferences\errmessage{I was expecting to see
\noexpand\endreferences before now;  I have inserted it here.}}
\let\r@ferences=\references
\def\references{\r@ferences\def\endmode{\r@ferr\par\endgroup}}

\let\endr@ferences=\endreferences
\def\endreferences{\r@fcurr=0
  {\loop\ifnum\r@fcurr<\r@fcount
    \advance\r@fcurr by 1\relax\expandafter\r@fis\expandafter{\number\r@fcurr}%
    \csname r@ftext\number\r@fcurr\endcsname%
  \repeat}\gdef\r@ferr{}\endr@ferences}


\let\r@fend=\endpaper\gdef\endpaper{\ifr@ffile
\immediate\write16{Cross References written on []\jobname.REF.}\fi\r@fend}

\catcode`@=12

\def\reftorange#1#2#3{$^{\cite{#1}-\setbox0=\hbox{\cite{#2}}\cite{#3}}$}

\citeall\refto          
\citeall\ref            %
\citeall\Ref            %
%
\gdef\journal#1, #2, #3, 1#4#5#6{{\sl #1 }{\bf #2}, #3 (1#4#5#6)}
\def\same#1, #2, 1#3#4#5{{\bf #1}, #2 (1#3#4#5)}

\def\pra{\journal Phys. Rev. A, }
\def\prb{\journal Phys. Rev. B, }

\def\prl{\journal Phys. Rev. Lett., }

\def\rmp{\journal Rev. Mod. Phys., }

\def\jpa{\journal J. Phys. A, }

\def\jsp{\journal J. Stat. Phys., }

%
\input jnl
\tolerance=500
\def\today{\ifcase\month\or January\or February\or March\or April\or
  May\or June\or July\or August\or September\or October\or November\or
  December\fi \space\number\day, \number\year}
\def\references{\head{References}\beginparmode\frenchspacing
  \singlespace\parindent=0pt \leftskip=1truecm \parskip=8pt plus 3pt
  \everypar{\hangindent=\parindent}}
\input reforder
\catcode`@=11
\def\refis#1#2#3\par{\ifuncit@d{#1}\w@rnwrite{Reference #1 uncited.}%
  \else\expandafter\gdef\csname r@ftext\csname r@fnum#1\endcsname
  \endcsname%
  {\writer@f#1>>#2#3\par}\fi}
\catcode`@=12
\input journals
\rightline{\today}

\def\b{\beta}

\def\tobs{\tau_{\rm obs}}

\def\terg{\tau_{\rm ergodic}}
\def\tout{\tau_{\rm outlet}}

\title
A New Look at Broken Ergodicity
\vskip .25in

\author D.L. Stein
\affil
Department of Physics
University of Arizona
Tucson, AZ 85721

\vskip .25in

\author  C.M. Newman
\affil
Courant Institute of Mathematical Sciences
New York University
New York, NY 10012
\vskip .25in
\abstract

We study the nature and mechanisms of broken ergodicity (BE) in specific random
walk models
corresponding to diffusion on random potential surfaces, in both one and high
dimension.  Using
both rigorous results and nonrigorous methods, we confirm several aspects of
the standard BE
picture and show that others apply in one dimension, but need to be modified in
higher
dimensions.  These latter aspects include the notions that at fixed temperature
confining
barriers increase logarithmically with time,  that ``components'' are
necessarily bounded regions
of state space which depend on the observational timescale, and that the system
continually
revisits previously traversed regions of state space.  We examine our results
in the context of
several experiments, and discuss some implications of our results for the
dynamics of disordered
and/or complex systems.

\endtitlepage

\body

\noindent 1.  $\underline {\rm Introduction}$.

\medskip

When a system has many metastable states, it may become trapped for long times
in some subset
of its total state space, making it difficult to compare experimental results
with
calculations based on the usual Gibbs formalism.   A viewpoint commonly called
``broken
ergodicity'' (BE) has evolved to serve as a qualitative guide for the
understanding of some of
the dynamical and thermal properties of these systems.  This has been extremely
useful in
several respects, but we are still hampered by the lack of a real theory.

Much of the problem is the difficulty of characterizing the nature of
metastability in real
systems.  As a result, the standard picture of BE which has emerged (to be
described below) is
based largely on intuition and simple pictures of what these state spaces may
look like.  All
basically involve diffusion of a particle (the system) on a rugged landscape,
which may or
may not possess correlations.  These pictures can all be described as diffusion
in a strongly
inhomogeneous environment.

While many of the results obtained in this way are compelling (and, as we will
discuss
below, almost certainly correct in a wide variety of situations), progress has
been slow, at
least partially because of the lack of specific models to test these ideas on.
In this paper
we will attempt to do just that; we will examine some simple, well-known models
and see how
broken ergodicity arises in them.
These models are representative of uncorrelated random potentials.  Reasoning
based on
random walks on such potentials has guided much of the thinking about how BE
operates in
disordered systems.\refto{models}  We will not address in this paper the
question of the
accuracy of such assumptions; \ie whether random walks on rugged landscapes are
useful for
modelling dynamics of some disordered systems.   Our only goal here is to
introduce clear,
well-defined models and to study their long-time behavior in the strongly
inhomogeneous limit.

The analysis will be based on some rigorous results obtained in an earlier
paper,\refto{NS1}
hereafter referred to as I, and some nonrigorous results obtained in
another,\refto{NS2}
hereafter referred to as II.  We will find that while many of the central ideas
of standard BE
apply to these models, there are some surprising deviations from important
elements of the
conventional picture.  We will find that this is at least partially due to the
fact that
while all workers in the field recognize that the relevant state spaces for
physical systems are
high-dimensional, much of the intuition about BE is nevertheless based on what
are ultimately
one-dimensional pictures.  We will see explicitly how the presence of many
dimensions
considerably changes the standard analysis.

The paper is organized as follows:  In Section 2, we review some of the basic
features of BE
pertinent to the analysis contained below.  In Section 3, we introduce two
simple models
of a random walk in a random environment (RWRE), and review our earlier results
within the
context of these models.  In Section 4, we analyze the behavior of broken
ergodicity in these
models, in both one and high dimensions.  In Section 5, we discuss and
summarize these
results, and make a few brief remarks about experiment.

\medskip

\noindent 2.  $\underline {\rm Broken\ Ergodicity}$.

\medskip

Because the phenomenon of broken ergodicity has been discussed at great
length in the literature, we here review only those aspects of it which are
relevant for the
cases under consideration.  The importance of non-ergodicity in disordered
systems,
particularly spin glasses, was emphasized early on by Anderson\refto
{PWA1,PWA2}.  Early analyses
and applications were given by J\"ackle,\refto{Jackle} Palmer\refto{Palmer1,
Palmer2}, and van
Enter and van Hemmen\refto{vEvH}.  The presentation by Palmer is especially
comprehensive and
accessible; most of the discussion in this section follows his treatment.  We
are concerned here
only with some of the central ideas of BE; for a complete overview, we refer
the reader to
the above papers.

We are primarily interested in cases where ergodicity is broken because the
observational
timescale ($\tobs$) falls within a continuum of relaxational or equilibrational
timescales
intrinsic to the system, as is commonly believed to occur for glasses and spin
glasses.
(See \Ref{Palmer1} and \Ref{vEvH} for other examples, including the more
familiar situation
of broken symmetry.)  This may occur either in the presence or absence of a
phase
transition.  The former is typically indicated by the state space breaking up
into two or
more disjoint components, separated by free energy barriers which diverge in
the
thermodynamic limit.  Broken ergodicity can and does occur, however, when the
system possesses {\it metastable\/} states surrounded by {\it finite\/} free
energy barriers.
Because the typical timescale for escape from a metastable state grows
exponentially with the
barrier, these need not be large for ergodicity to be broken on laboratory
timescales.

The central idea is that state space can be decomposed into {\it components\/}
which
are not necessarily intrinsic to the system, but rather depend on the
timescale.  Components
are defined by the probability of confinement on some timescale $\tau$:  if the
system is in a given
component at time $0$, then the probability that it has not escaped from the
component by time $\tau$ is greater than some specified (fixed)
probability $p_0$.\refto{Palmer1}  Clearly, the definition of component depends
both on the
specified timescale $\tau$ and the probability $p_0$.  It is also assumed that
on the same
timescale, the system is ergodic {\it within\/} the component; \ie the system
visits a
representative sampling of states within the component so that the state space
average
equals the time average, so long as one confines the averaging to states within
the
component.\refto{Palmer1}

What is the confinement mechanism?  We are interested here in {\it
structural\/} confinement
mechanisms rather than dynamical (\ie the existence of possible constants of
the motion.)
In the former case, the system is confined to a component because the smallest
free energy
barrier that must be surmounted in order to escape corresponds to an escape
time large
compared to the observational time.  The standard picture envisions a very
mountainous
terrain, with a series of isolated lakes and puddles in various valleys.  The
``water level'' corresponds to the largest free energy scale which
the system can sample on a given temperature and time scale.  If temperature is
held fixed, and time is allowed to increase, the water level steadily rises,
with lakes
merging into bigger lakes and bigger lakes into oceans, leading to a
hierarchical merging of
components\refto{Krey, Palmer1, Palmer2, PS, Sibani}; see Fig.~1.  (One can
arrive at the same
picture by fixing time $t$ and letting temperature $T$ increase;  the height of
the water
level scales as $T\log t$.)

Until the system surmounts the highest barriers, ergodicity remains broken; as
soon as the
system surmounts some free energy barrier, it finds itself in a larger
component which is
confined by higher free energy barriers.\refto{Palmer1, Palmer2, PS}  So the
confining free
energy barriers which the system must surmount, at fixed temperature, increase
logarithmically with the time.  Also, because the system is now ergodic within
the larger
component, it will continually revisit the previous smaller one, which is now a
subset of
the portion of state space which it currently explores.

We will now examine these {\it ans\"atze\/} in two specific
models, both of which possess a continuum of free energy barriers --- and
therefore a
continuum of relaxational timescales.  We will find that the above
picture needs to be modified in several respects:  while it
precisely describes a {\it one-\/}dimensional version of our models,
there are important differences in higher dimensions.

These high-dimensional models are indeed the relevant ones, since within the
context of BE
one is usually referring to the evolution of a system in some high-dimensional
state
space.  As in I, one often models this state space as some graph ${\cal G}$.
The vertices of the graph correspond to the states themselves, and the
edges correspond to transitional paths between pairs of states.  The
dimensionality of the graph scales with $N$, the number of degrees of
freedom in the system.  We will follow this general procedure here.

We begin with a description of our dynamical models, which are just
specific examples of a much studied process -- the random walk in a random
environment.\reftorange{RWRE}{Sinai, PV}{DL}

\medskip

\noindent 3.  $\underline {\rm Inhomogeneous\ Random\ Walk\ as\ Invasion\
Percolation}$.

\medskip

In this section we review the results of earlier work.  We consider two
different dynamical
models within the overall context of the RWRE:

1)  Model A --- ``edge'' model.  Here we consider a graph ${\cal G}$ in which
the sites correspond
to states and the (nondirected) edges to the dynamical pathways which connect
them.  For
specificity, we take ${\cal G}$ to be the lattice $Z^d$, although this is
unnecessary for our
results.\refto{hypercube}  We assign nonnegative, independent, identically
distributed random
variables to each of the edges; these represent the energy barriers which must
be surmounted in order
to travel between pairs of sites connected via the edges. (We assume that the
distribution of
these variables is continuous so that all barriers have distinct energies.)
If $W_{xy}=W_{yx}$ is the value assigned the edge
connecting sites $x$ and $y$, then the rate to travel {\it in either
direction\/} between
$x$ and $y$ is taken to scale with inverse temperature $\b$ as
$$
r_{xy}(\beta)\sim\exp[-\b W_{xy}]\quad . \eqno(3.1)
$$

2)  Model B --- ``site'' model.  Here we assign random variables to {\it
both\/} sites
and edges of ${\cal G}=Z^d$.  Because we wish to view each site as
corresponding to a locally stable
state (\ie as the minimum energy configuration within a ``valley''), and each
edge as
again corresponding to an energy barrier, the values assigned to sites and
edges cannot be
identically distributed --- both sites touching an edge must have lower
assigned energy
values than that of the edge itself.  (As in model A, we simplify matters by
choosing each distinct edge
to connect a single pair of sites.)  A simple way to implement this
is to choose the site variables independently from a single negative
distribution, and the
edge variables independently from a single positive distribution.  However, if
the model is to correspond to any sort of physical (random) potential, the
distribution for the site variables, whatever its form, {\it must be bounded
from below.\/}
This is relevant to the analysis given below.  The distribution for the edge
variables, of course, need not be bounded from above.

If $W_x$ is the variable assigned to site $x$, then the equilibrium
probability density over sites $\pi_x(\b)$ scales with $\b$ as   $$
\pi_x(\b)\sim\exp[-\b
W_x]\quad . \eqno(3.2) $$ Detailed balance then requires that
$$
\pi_x(\beta)r_{xy}(\beta)=\pi_y(\beta)r_{yx}(\beta)\quad ,
\eqno(3.3)
$$
where $r_{xy}$ is understood as the rate to go from $x$ to $y$.  The rates,
satisfying \eq{3.3}, are chosen so that
$$
r_{xy}(\beta)\sim\exp\left[-\beta(W_{xy}-W_x)\right]  \eqno(3.4)
$$
and
$$
r_{yx}(\beta)\sim\exp\left[-\beta(W_{xy}-W_y)\right]\quad .
\eqno(3.5)
$$

We note that for detailed balance to hold in model A, the probability density
over sites
must be site-independent; that is, model A corresponds to model B
with all energy minima degenerate.  Despite the extreme simplicity of model A,
it has a rich
and suggestive dynamical behavior, and we therefore include it in our analysis.

Our main result in I is a rigorous statement about the {\it order in which
sites are visited
for the first time\/}, given an arbitrary starting site.  We will see later
that it is easily
extended to a result about which sites are visited (or not visited) on a given
timescale, and
the nature of that process, which is of central interest in a BE treatment.

The assignment of random variables in both models defines an ordering on the
(undirected) edges of ${\cal G}$ in
which $\{x,y\}<\{x',y'\}$ if $W_{xy}<W_{x'y'}$.  That is, the barriers
are ordered by increasing height.

Our theorem in I, which applies to both Model A and Model B, states the
following:  for any
configuration of the $W_{xy}$'s (and $W_x$'s), as
$\b\to\infty$, the sequence in which sites are visited, starting from some
arbitrary initial site $x_0$, converges to the invasion percolation
sequence with the same initial site and the same edge ordering.  (For a precise
statement of
the theorem and its proof, see I.  For a brief review of invasion percolation,
see
Appendix A below.)

Intuitively, the above result is quite reasonable.  As the temperature is
lowered, the
timescales for transitions to neighboring sites diverge from one another.  With
increasingly
high probability, the system will make a transition over the lowest barrier
available to
it.  Although the idea is quite simple, it has important, and heretofore
unappreciated,
consequences, due to the geometry and structure of invasion percolation.  These
will mostly
be discussed in the next section. First, we quote one more result which will be
central to
our later discussion.  This is a nonrigorous result about the global
connectivity
structure of invasion percolation, discussed by the authors in II.  It says the
following:\refto{NS2}

For invasion percolation on the lattice $Z^d$ when $d<8$, there is an
essentially {\it unique\/} invasion region.  That is, given any two starting
sites, their invasion regions will be the same except for finitely many
sites (with probability one).  However, when $d>8$, there are {\it
infinitely\/} many disjoint invasion regions.  That is, given two starting
sites far from each other, their invasion regions will totally miss each
other with high probability.  The asymptotic dimension of these invasion
regions in high $d$ is four.

There is a picturesque way to view this: Let $p_c$ denote the critical value
for independent bond percolation on $Z^d$ and let $w_c$ denote the energy level
such that Prob$\ (W_{xy}\le w_c)=p_c$.  Then from any starting site the
invasion
process, as time increases, focuses on the so-called incipient infinite
cluster at $p_c$ in the corresponding independent bond
percolation problem.  Once the process finds an infinite cluster of edges
 with energy levels $\le w_0$, where $w_0>w_c$, it never
again crosses an edge with $W_{xy}>w_0$.  One can then consider the invasion
process from any point as following a ``path'' which eventually leads to
``the sea'' at infinity.  In less than eight dimensions, all invasion regions
from different points eventually
follow the same path to the sea.  Along the way, all individual paths merge,
some sooner,
some later.

However, in greater than eight dimensions, there are an infinite number of
disjoint paths to infinity.  Indeed one should think of infinitely many
distinct seas, each of which has many tributaries,
\ie invasion regions which flow into it, or equivalently, an (infinite) set of
sites whose
invasion regions connect to it.  The process flows into one of these seas,
and {\it it will never visit any sites which connect, via the invasion process,
to any of the
other seas.\/}  Because models such as A and B are often used to (abstractly)
describe state
spaces in very high dimensions, this picture is a crucial component in what
follows.

We close this section with an important remark about the theorem from I
described above.  It is
well known that the RWRE asymptotically approaches
ordinary diffusion at long times.\refto{PV}  (This is also the case above two
dimensions for RWRE's which, unlike those treated here, do not
satisfy detailed balance\refto{DL}. It need not be the case in such models
for one dimension or in detailed balance models with sufficiently correlated
environments\refto{Sinai}.)  But our picture seems to contradict the diffusion
picture.  In fact,
both are consistent, because each corresponds to a different method of taking
the limits
time$\to\infty$ and $\beta\to\infty$, and the behavior of each model is
sensitive to this.

Previous treatments\refto{PV, DL} studied the case where temperature is fixed
and time
goes to infinity; in that case, the RWRE will exhibit normal diffusive
behavior.
In our picture, we first focus on a particular site $y_0$.  Suppose that $y_0$
is the $157^{\rm th}$ site invaded by the invasion process described earlier.
Our theorem states that, as $\b\to\infty$, the probability that $y_0$ is also
the
$157^{\rm th}$ site visited by the RWRE converges to one.
This implies that there exists a temperature-dependent
timescale --- an ergodic time, so to speak --- beyond which our picture breaks
down and
normal diffusion takes over (or equivalently, ergodicity is
restored).\refto{erg}  The ergodic time
diverges as temperature goes to zero.  A timescale of this type is
a common feature in most systems which break ergodicity.  We will discuss this
further in the following
sections, but meanwhile note the rigorous illustration, in a specific model, of
an important
feature\refto{Palmer1} of broken ergodicity --- the way in which limits are
taken is crucial!

\medskip

\noindent 4.  $\underline {\rm Broken\ Ergodicity\ in\ the\ RWRE}$.

\medskip

\noindent A.  {\it One-Dimensional Picture}

We first consider the RWRE in one dimension.  It will be sufficient to
consider only Model A in this case, because here there is no significant
qualitative difference between the two models.  This is not quite true
in higher dimensions.

For specificity, let the edge random variables, which correspond to
barriers, be chosen independently from the positive half of a Gaussian
distribution with mean zero and variance one.  Consider
the behavior of the diffusing particle for large time and low temperature.  It
is easy to see
that in this case, all of the assertions made in Section 2 are correct
after some initial transient time (the larger $\b$ is, the shorter this
transient time becomes).  On some observational timescale $\tobs$, the
particle is trapped with high probability between two barriers, neither of
which are surmountable (with some prespecified probability) on a timescale
of the order of $\tobs$.  If one is willing to wait considerably longer (on
a logarithmic timescale), then the length of the line segment the particle
explores is correspondingly larger, surrounded at each end by suitably
large barriers.  It is not hard to show that these grow in the manner
specifed in \Ref{Palmer1} and \Ref{PS}:  $\Delta F_{\rm esc}\sim\log\tobs$.

If one were to watch a greatly speeded up movie
of the particle motion, it would look something like the following.  After
diffusing to the right (say) some distance, the particle encounters a
barrier significantly larger (compared to $1/\b$) than any it has
previously crossed.  The particle is effectively reflected to the left,
where it undergoes a net diffusive motion until it encounters a new
barrier significantly larger than any previous ones, including the
original reflecting barrier.  (Prior to this, however, it may have
encountered barriers smaller than the first reflecting barrier but larger
than any others and subsequently have bounced back and forth a number of
times.)  The particle ``reflects'' off this barrier and begins a net
diffusive motion to the right.  Eventually, well to the right of the first
reflecting barrier, it encounters a new barrier of yet greater magnitude
than any previous ones, which reflects it back to the left, and so on.
Informally, the process resembles a game of diffusive ping-pong with
asymmetrically receding paddles.

\medskip

In two and higher dimensions, the picture changes dramatically.  We will
see that in both models A and B, several of the standard BE assumptions break
down.
Among the most important of these is that as time increases, while
components grow larger, {\it they do not contain previously visited
portions of state space}.  Perhaps more surprisingly, in Model A the
confining barriers (\ie outlets --- see below) do not increase with time; they
instead {\it
decrease}, asymptotically approaching a constant from above.  In Model B, a
constant
barrier value is also approached (although not necessarily
monotonically).\refto{Note}  In neither case do the barriers grow
logarithmically with time.  In order to see where these surprising features
come from, we
return to a more extensive discussion of invasion percolation before examining
the models.

\medskip

\noindent B.  {\it Ponds and Outlets}

\medskip

We briefly digress from our discussion of broken ergodicity in the RWRE
to examine the process whereby invasion percolation ``finds a path'' to
infinity.  In accordance with our theorem proved in I, this will be
equivalent to the behavior of the diffusing particle in the RWRE under
an appropriate range of temperature and timescale.  The picture
presented in this section holds irrespective of whether there is one or
infinitely many disjoint invasion regions.

We first present the standard argument which connects the asymptotic
geometry of the invasion region to that of the incipient infinite
cluster at $p_c$ in the corresponding independent percolation problem.
(See Appendix A.)  We consider bond percolation on $Z^d$ in both cases.
Hereafter the term ``invasion region'' should be understood to mean
``invasion region starting from some arbitrarily chosen initial point
$x_0$''.  As in Appendix A, we can and do confine ourselves to the case
where the bond, or edge, variables are chosen independently from the uniform
distribution on $[0,1]$.

Given $x_0$, and a configuration of the bond variables, there exists a unique
invasion route to infinity.  Consider
all bonds whose values (\ie magnitudes of assigned random variables) are
smaller than some $p_1>p_c$.  By correspondence with the associated
independent bond percolation problem, these comprise a unique infinite
cluster, in addition to finite clusters of varying
sizes.\refto{Grimmett}  Therefore, once the invasion process reaches {\it
any\/} of the bonds within this infinite cluster, {\it it will never
again cross any bond whose value is greater than $p_1$\/}.

Consider next all bonds whose values are less than some $p_2$, where
$p_c<p_2<p_1$.  These too form a unique infinite cluster which is a
subset of the first, larger one.  When the invasion process reaches any
of the bonds within this newer infinite cluster, it will never again
cross any bonds greater than $p_2$.  It is easy to see that, as the
process continues, the invasion region will ``focus down'' to infinite
clusters of increasingly smaller maximum bond value, and will
asymptotically converge to the incipient infinite cluster of the
independent bond percolation problem at $p_c$.

It is important to note that the ``incipient infinite cluster'' is
not an infinite cluster at all (that is, there is no percolation at $p_c$),
but rather consists of a sequence of increasingly
large but disconnected clusters.  To visualize this, consider a
simulation of independent bond percolation when $p$, the probability that
a bond is occupied, equals $p_c$.  If one looks at a finite cube of
volume $L^d$, one will, when $L$ is large enough, indeed see that the
largest cluster of occupied bonds has linear extent of order $L$.
Suppose one now increases $L$ dramatically.  Again the largest cluster will
stretch
across much of the length of the box, but it may {\it not\/} contain the first
cluster, which in fact is finite (see Fig.~2).  At $p_c$, there are {\it no\/}
infinite clusters; nevertheless, any large but finite box
will have a cluster of comparable linear dimension.
This sequence of increasingly large but finite clusters may be thought
of as the incipient infinite cluster.  It should be noted that there are
alternate constructions\refto{Kesten, Chayes} which yield infinite clusters
different than, but closely related to, the above notion of incipient
infinite cluster.

Let us now examine more closely how the process of invasion occurs.
We utilize here a construction of Hammersley\refto{Hammersley} which,
although it predated invasion percolation, seems tailor-made for its analysis.
(Indeed, a modified construction can be used to analyze\refto{C2N} versions
of invasion percolation with trapping.)
Because $x_0$ is arbitrary, the process will generally invade some set
of relatively smaller-valued bonds before it has to invade a relatively
larger one to make its way toward infinity.
Picturesquely, the process is stranded on a pond, and has to invade
a relatively high outlet before it can escape.  The outlet
corresponds to the bond whose value is larger than that of all others
within the pond, but smaller than all others on the perimeter of the pond.  It
is
furthermore crucial to note two things:  first, that this first outlet will
be the {\it bond of largest value that the process will ever cross\/},
and second, that once this outlet is crossed, {\it the process will
not return to the first pond\/}\refto{recur}.  The significance of this ``diode
effect'' will be discussed below.

After crossing the first outlet, the process will find itself
on a second pond, and must invade an outlet of smaller value than the
first one.  In this way it invades a sequence of
successively smaller outlets (with bond values larger than but tending
toward $p_c$) on its way to the sea (see Fig.~3).
The general trend is for the ponds to grow successively larger, but this
need not be true monotonically.

Ponds and outlets can be defined precisely; we give two alternative definitions
here.  In the first, consider all possible paths to infinity from the starting
point $x_0$.  Each such path ${\cal P}$ will contain some bond of maximum
value; call it $b_{\cal P}$.  The first ``outlet'' is then the bond $b^*$
of {\it minimum\/} value from the set $\{b_{\cal P}\}$.  The first ``pond''
is the finite cluster connected to $x_0$ consisting of all bonds whose values
are strictly less than that of $b^*$.
The second pond and outlet can be found using the same procedure from the
starting point $x_1$,
where $x_1$ is the site which touches $b^*$ and is outside the
first pond.  This procedure can be repeated indefinitely to find ponds and
outlets of any order.

The second definition uses an alternative procedure.  Starting from $x_0$, one
considers the finite cluster connected to $x_0$ which consists of all bonds
with values less than $p=p_c$.  One then raises $p$ in a continuous manner,
causing the cluster connected to $x_0$ to grow.  At some sharp value of
$p$ (depending on $x_0$) the cluster becomes infinite; it is not hard to see
that
there will be a single bond connecting the (previously finite) cluster
containing
$x_0$ with infinity.  This bond is the first outlet, and all bonds in the
interior
finite cluster comprise the first pond.

We now return to the situation of a random walk in a strongly
inhomogeneous environment, and apply these ideas and results to see how
the process evolves.

\medskip

\noindent C.  {\it Time Evolution of a Random Walk in a Random Environment}

\medskip

We now examine these results within the context of broken ergodicity.  An
immediate
conclusion, which applies both to Models A and B, is that it is useful to
redefine the
notion of ``component''.   While the conventional definition (see Section 2)
can of course be
applied here, we propose an alternative characterization which we suggest may
be more useful
for Models A and B.  We propose here {\it two\/} kinds of components --- one
global and
one local, each of which reflects the natural structure of the models' dynamics
as uncovered by
our analysis.  The local type is similar to the usual component in several
respects, but
the global type is markedly different.  In both cases, however, there are
important differences
from conventional components, to be described below.  We begin by introducing
the notion of a
global component.

The global components correspond to the invasion regions.  Recall that below
eight dimensions, there exists a unique (asymptotic) invasion region, while
above eight, there
exist many.  Because any state space of interest will be high-dimensional, we
hereafter restrict
ourselves to this case.  Therefore, many disjoint invasion regions exist, and
which one a
diffusing ``particle'' (corresponding to the state of the system) finds itself
in depends on the
starting point (\ie the initial configuration of the system).  As long as the
time is less than
the ergodic time, defined above, the system is confined to a single one of
these components, and
will therefore not visit states corresponding to any of the others.  These
components differ in
two important ways from the usual kind of component:

\noindent $\bullet$  They are {\it intrinsic\/} to the system itself, and are
not defined
with respect to any observational timescale.  However, it is important to note
that, at
fixed temperature, there is a cutoff timescale above which the confinement
mechanism
breaks down.  There is no ``absolute'' confinement; all barriers are finite.

\noindent $\bullet$  They are {\it infinite\/} in extent.

We showed in II that the dimension of the invasion region is four in high
spatial
dimension.\refto{NS2}  This should be contrasted with the dimension of ordinary
Brownian
motion, which is two; as long as the confinement, or invasion, mechanism
dominates the
dynamics, the density of sites visited by the walker is higher than when
ordinary, free
diffusion takes over.  Some models of random walks in state
space\refto{Campbell} connect this
type of behavior with slower than exponential relaxation, but we will not
pursue the matter
here.

How long does the confinement mechanism hold?  Because the various transitions
occur on
exponentially different timescales, it is more fruitful to estimate the number
of sites visited
during the confinement period, rather than the actual time.  A very rough
argument suggests that the former will scale as
$e^{a\beta}$, where $a$ is a constant depending on the model chosen and form of
the
distribution.  We will not determine $a$ below --- the exponential dependence
on $\beta$ is the relevant conclusion.

Perhaps what is most surprising about the above picture is that, as long as the
confinement mechanism holds and the system breaks ergodicity, the evolution of
the system
(in terms of states visited) is largely deterministic, depending only on the
starting
point.  Of course, the particle will still diffuse among the states allowed by
the above
confinement mechanism, but it does so in a manner which again differs
considerably from
all previous pictures of which we are aware.  In order to see how this occurs,
we turn to
a discussion of each of our specific models.

\smallskip

{\it Model A\/}

\smallskip

We have already seen how the idea of components, as determined by observational
timescales, must be modified in these models; they are replaced by the notion
of disjoint
invasion percolation regions.  But within each of these global components, we
retain
considerable structure in the form of ponds and outlets.  These play an
important role in
the time evolution of the system, and correspond more closely to the
conventional idea of
components.  However, some important differences exist here also:

\noindent 1)  In conventional BE, the system, as it diffuses in state space,
returns infinitely often to the same region of state space within which it was
confined
at earlier times.  The opposite is true here --- once the system leaves a pond,
{\it it never returns.\/}  There is a type of ``diode effect'', where, on
the scale of regions the size of ponds, the walker always moves forward,
never backward (see Fig.~3).  Some possible experimental manifestations of this
will be
discussed in Section 4.

This diode effect is quite different from the nonreturn of ordinary random
walks.  For example,
in $d=2$, the diode effect remains valid even though ordinary random walks are
recurrent.  For
large $d$, the invasion region dimension is {\it twice\/} that of an ordinary
random walk, as
mentioned above.

\noindent 2)  In conventional BE, the system must surmount increasingly high
barriers as
time progresses.  The opposite again holds here --- the barriers which confine
the
system (\ie the outlets) diminish steadily as time increases.  The landscape
through which the
diffusing system travels becomes increasingly flat. (The particle often sees
many high peaks in
its vicinity, but it avoids them.)  This is true in all dimensions greater than
one.

In regard to (2), it is important to consider the following point.  Because
entropy (and
presumably entropy barriers) plays an important role in the physics of
disordered systems, it
might be argued that there is still a large entropy barrier to find the outlet
which leads off
a pond, and that this barrier increases with time.  While the typical size of
ponds undoubtedly
increases with time, it does {\it not\/} appear to be the case that entropy
barriers are
playing a significant role.

In fact, consider the system immediately after it has escaped the $n^{\rm th}$
pond.  We argue here that most of the time expended up to that point was used
in getting {\it to\/} the
pond in the first place.  Finding the outlet on any pond is not a
needle-in-the-haystack problem; the system is not wandering around aimlessly in
state
space, eventually finding the outlet by remote chance.  The dynamics of the
invasion
percolation mechanism restrict the system to a particular pond at any time, and
the
timescale for confinement within that pond is simply $\exp [\beta W_{n}^*]$,
where
$W_{n}^*$ is the value assigned to the outlet for that pond.
During this time the process thermalizes within the pond.

For entropy effects to counteract the decreasing values of $W_{n}^*$ as $n$
increases,
it would seem to require exponentially increasing pond size, whereas pond
size almost certainly increases much more moderately, probably as a power law.
We conclude that for Model A (and, as we'll see, for Model B also) the picture
shown in
Fig.~1 is a purely one-dimensional picture.  In any higher dimension
(irrespective of
whether there are one or many global components), the ``water level'' does {\it
not\/}
rise as time increases at fixed temperature (or temperature increases at fixed
time).
Viewed from $x_0$, the water level initially rises, {\it but then stays forever
fixed\/}, because it finds a path
to the ``sea'' (\ie to infinity), into which it empties.  Any additional water
poured in
simply escapes to infinity.

\smallskip

{\it Model B\/}

\smallskip

Model B is more satisfactory than A in that it corresponds to a system with
nondegenerate states.  However, the only difference with Model A in the BE
context is in
point (2) above.  Here again, the height of the confining barrier, or outlet,
is declining
towards a limit $w_c$, corresponding to $p_c$ in the independent percolation
model.
However, the energy of the lowest ``valley'' within a pond generally becomes
more
negative as time progresses (and the system explores larger ponds),
asymptotically (but
slowly) approaching the minimum of the distribution; call it $w_{min}$.
Therefore, the barriers which the system must surmount to escape successive
ponds
asymptotically approach $w_c-w_{min}$.

\medskip

\noindent 5.  $\underline {\rm Discussion\ and\ Conclusions}$.

\medskip

\noindent {\it Hierarchies}

\smallskip

In all cases most of the predictions which follow from a BE viewpoint remain
valid.  For example, the idea of components nicely describes experiments on
spin
glasses\reftorange{TT}{NKH}{Chamberlin} wherein the system displays reversible
behavior when
temperature is first lowered, then raised, but irreversible behavior when it is
first raised,
then lowered.  (See also \Ref{Refrigier}, which describes a similar effect
within the context
of ageing of spin glasses.)  That can be viewed within the context of
hierarchically nested
components.\refto{Krey, Palmer1, Refrigier, Stein87, Sibani}  However, this
irreversibility
signature easily arises in our models also, and in fact it is a consequence of
a very wide
variety of models with inhomogeneous energy landscapes in state space.

To see how it arises in our picture, simply consider some timescale at which
the system can be found on a particular pond.  If the temperature is lowered,
the
system remains confined to the pond (on the same timescale).  When the
temperature is then raised, it merely restores the original situation, so the
observed
behavior appears reversible.  If this procedure is done in the opposite order,
however,
the system can diffuse quickly to a different pond (or even to a different
global
component altogether, if the temperature is raised sufficiently).  Lowering the
temperature
leaves the system in a different region of state space, leading to the
observation of
irreversibility.

There is little need of global hierarchical behavior in our picture.  Locally,
one
might imagine there is some, in that one can think of confinement within a
subset of a
pond upon lowering temperature at a fixed timescale, at least
within a limited range of temperature and timescale.  However, the more global
situation
is not hierarchical in any strong sense, neither in terms of the
``pond-hopping''
scenario nor in terms of the global components.

\smallskip

\noindent {\it Brief Remarks on Some Selected Experiments}

\smallskip

There may be an experimental way of distinguishing between the conventional,
hierarchically nested component picture, and the outlet-and-pond picture for a
particular system.  Suppose that the temperature is raised by only a small
amount.  It is
reasonable to expect that the system cannot diffuse far on relatively short
timescales,
so that if the temperature is then restored and the experimenter waits, the
system may
eventually display earlier behavior, particularly if a hierarchical picture
applies.  (In terms
of going up or down a hierarchical tree, this would be equivalent to going up
only one or two
levels; so after some waiting period, the system has some reasonable
probability of
rediscovering its original state.)  This probability would be far smaller if
the system is
diffusing from pond to pond, with little prospect of returning to those visited
earlier.

An experiment in a similar spirit was performed on Ag:Mn spin
glasses,\refto{Chamberlin} where
magnetization of zero-field-cooled samples was measured after application of an
external field.
Turning on a magnetic field was assumed to ``randomize'' the energy surface; in
the context
of our Model B, for example, it might correspond to reassigning values to the
site and bond
variables, thereby beginning a new diffusion process.\refto{Narayan}

It was hypothesized that the change in the energy surface with field would take
place
continuously, so that if the external dc field were changed by a very small
amount, a reversible
change in magnetization would also be seen.  This was not observed, however,
even for the
smallest applied fields ($\sim 40$ mOe); the magnetization always displayed
an irreversible drift governed by a characteristic
quasilogarithmic time dependence.\refto{TT,Guy,KK, Chamberlin}  The explanation
given was that any
field, no matter how small, completely ``scrambles'' the energy surface; but a
simpler
explanation might be that the surface is largely unchanged, and one is simply
observing the
diode effect discussed earlier.\refto{FH}

It is also of interest to note that in the irreversibility signature of
zero-field-cooled
spin glasses discussed in several papers,\reftorange{TT}{NKH}{Chamberlin} the
magnetization
appears to remain {\it constant} when temperature is first lowered, then
raised, in nonzero
field.  This doesn't seem consistent with Fig.~1, where if the system is at
temperature
$T_B$, the magnetization is an average over the states contained within $B$.
Upon lowering
the temperature to $T_A$, which confines the system to the smaller region $A$
of state space,
it seems reasonable to expect the magnetization to change; but this is not
observed.
Although at first glance the same arguments might seem to imply a change in
magnetization
in the invasion picture, due to confinement to a subset of a pond upon lowering
of the temperature, we suggest that this is not the case.  This is simply
a consequence of the fact that the pond size is slowly growing, so that a
typical pond is small enough to preclude a large variation in the macroscopic
characteristics of its subcomponents; in contrast, components in the standard
BE picture (while finite) are unrestricted in size.

\smallskip

\noindent{\it Restoration of Ergodicity\/}

\smallskip

Our result, that the dynamics of many RWRE models follow that of invasion
percolation,
is sensitive to the manner in which the limits $T\to 0$ and $t\to\infty$ are
taken.  The result is
rigorous (for A, B, and presumably related models) when time diverges
appropriately as
temperature goes to zero.\refto{NS1}  It is also rigorously known that, when
time goes to infinity for fixed temperature,
the RWRE tends toward ordinary diffusion (in the sense that the mean square
displacement scales
linearly with time).\refto{PV}  Either situation can be (and
often is) realized experimentally, but because our result breaks down in the
second case, we
examine this situation a little more closely here.

How should we expect these models to behave when temperature is fixed (at some
value small
compared to the majority of barrier heights) and time increases?  As discussed
earlier, there
will be an ergodic timescale at which the system is likely to escape the global
component
(invasion region) in which it finds itself.  Beyond this timescale we expect
ordinary
diffusive behavior, but of a rather funny sort:  the system will mostly hop
from component to
component on the ergodic timescale, but after it finds itself in a new
component, it stays
there roughly for another ergodic time.  So on shorter timescales, one will
find the
system within a global component; on longer timescales, it diffuses between
components.
Unless one is examining the system on timescales extremely large compared to
$\terg$ (often
well beyond the reach of laboratory timescales), one will observe the basic
picture
described above.

One can also conceive of a ``pre-ergodic'' timescale, beyond which the system
remains within
its initial global component, but skips some ponds and/or reshuffles the order
of pond-hopping inside that
component.  It's not clear whether this timescale is very different from
$\terg$, considering
that a nonnegligible fraction of barriers which confine the system to a global
component are not significantly
larger than those within the component itself.  One can envision other models,
however,
where these timescales may be considerably far apart.

\smallskip

\noindent{\it Further Remarks on Timescales\/}

\smallskip

In addition to $\terg$, an important role is played by the timescale $\tout$
when the
{\it first\/} outlet is reached. Let $W^*$ denote the height of this ``highest
barrier to the sea''. In Model A, for temperatures $T\ll W^*$, $\tout$ scales
like $\exp [\beta W^*]$. For $t\gg\terg$, diffusion has taken over (with
ergodicity restored) and the invasion picture is not valid, while for
$t\ll\tout$ the invasion picture is essentially the same as that of
conventional BE.
The novelty of the invasion picture (with its diode effect, decreasing barriers
to the sea, \etc)
is thus restricted to times between $\tout$ and $\terg$.

Unlike the case of $\terg$, the nature of the scaling constant $W^*$ for
$\tout$ is known exactly\refto{Hammersley}:  For a given starting point $x_0$,
$W^*$ depends on the configuration of $\{W_{xy}\}$. Its distribution (inherited
from that of $\{W_{xy}\}$) is given by the simple formula,
$$
{\rm Prob\ }(W^*\le p) = \theta_{d}(p), \eqno(5.1)
$$
where $\theta_{d}(p)$ is the usual order parameter (\ie the percolating
network density) for independent bond percolation on $Z^d$ with bond density
$p$. In high dimensions, $W^*$ would typically be of the same order as the
critical value $p_c$ (which is $1/(2d-1) + O(1/d^2)$).

Of course, the invasion picture is exact only in the limit $T\to 0$.\refto{NS1}
For fixed small $T$, the random walk order of visitation may differ from the
invasion order for pairs of bonds whose $W_{xy}$-values differ by $O(T)$.
Nevertheless, the global structure of the invasion should be matched for
$t$ below $\terg$.

\smallskip

\noindent{\it How General Is This Picture?\/}

\smallskip

While there is a good deal of indirect evidence that energy surfaces in
glasses, spin glasses,
and other disordered systems are rugged (in the sense of having many metastable
states of
varying depths surrounded by barriers of varying heights)\refto{BY} there is
very little
firm knowledge about the detailed structure of these landscapes.  Certainly the
uncorrelated
landscapes discussed here are too simple; the real issue is how much our
picture is altered by
correlations in realistic landscapes (see also Footnote 17).  We cannot answer
this question
definitively.  However, it is suggestive that aspects of the conventional
picture emerge in the
one-dimensional case and disappear in higher dimensions, and we are willing to
speculate that the
essential features we discuss above are more robust than our simple models may
indicate.

The above analysis implies that much of the intuition utilized in BE studies of
disordered
systems is based on a strictly one-dimensional picture that recurs throughout
the literature.  This is
the well-known diagram of a rough surface as a function of some abstract
configurational
coordinate. (See Fig.~1; also Fig.~2 in \Ref{Stein88}.)  While all workers in
this subject are
well aware that the real picture is many-dimensional, that realization has
never, to our
knowledge, been effectively utilized.

It seems reasonable to conjecture that the basic picture of invasion
percolation described in
this paper will continue to hold, perhaps in a modified form, in more realistic
models of
actual systems.  We expect in particular that our observation that high
barriers will play
little role in confining the system will hold in almost any model, for a simple
reason ---
while they must be surmounted in one dimension, they can easily be gotten
around in high
dimensions.  In any case, one should be aware of alternative viewpoints and
pictures, such as
those described above, and be prepared to think about experimental and
numerical results in
these or other alternative frameworks.

\smallskip

\noindent{\it Summary\/}

\smallskip

We have presented a picture of broken ergodicity based on an analysis of
specific models.
Although the models are simple,  the analysis leads to several surprising and
novel
conclusions.  These include:

\noindent$\bullet$  There is a natural mechanism of ergodicity breaking
determined by the system itself and
independent of the observational timescale.  That is, the global components
(and also the ponds within them) can in principle be determined
independently of observational timescale.  This gives rise to an intrinsic
ergodic timescale,
below which ergodicity is broken and above which it is restored.

\noindent$\bullet$  Within each global component, a ``pond and outlet'' picture
provides a framework for interpreting traditional broken ergodicity.
In particular, the traditional picture is valid within each pond, where local
components can be defined with respect to observational timescale in the
usual way.

\noindent$\bullet$  Global components are infinite in extent.  Moreover, our
picture requires
an important alteration of the usual image of confining barriers growing
proportional to the logarithm of the observational timescale.  In our models,
confining
barriers in one case {\it decrease\/} with time in a natural way, and in both
cases
asymptotically approach a constant value.

\noindent$\bullet$  Once it surmounts a confining barrier
(an ``outlet''), the system does {\it
not\/} return to the portion of state space previously explored.  Over long
times, there is progressive motion away from the starting point, replacing
the traditional version of a growing, diffusively explored region in which
the system re-explores earlier configurations infnitely often as time goes to
infinity.

\medskip

Why is it important to study these systems from this viewpoint?  As Palmer
correctly points
out,\refto{Stein88} we cannot apply statistical mechanics blindly to these
systems until we
characterize the broken ergodicity, and in particular it is necessary to
determine the
component structure.  This last problem has remained in a primitive state, and
has only
infrequently been tested against actual models which can be thoroughly
analyzed.  Our hope is
that the present analysis will lead to treatments of increasingly complex
models, with a
continual refining of our understanding of how it is that real systems break
ergodicity.

\bigskip

\noindent $\underline{\rm Acknowledgments}$.

\medskip

This research was partially supported by NSF Grant DMS-9209053 (CMN) and by DOE
Grant
DE-FG03-93ER25155 (DLS).  DLS acknowledges the hospitality of the Courant
Institute of Mathematical Sciences of New York University during a sabbatical
stay, where
this work was begun, and the Aspen Center for Physics, where part of this
research was carried
out.  DLS also thanks Richard~Palmer for introducing him to the subject of
broken ergodicity and
convincing him of its importance.  Both authors are grateful to Walter~Nadler
for useful suggestions
on the manuscript.

\vfill\eject

\noindent $\underline{\rm Appendix\ A}$.

Invasion percolation\refto{inv} can be defined as a process either on edges
or on sites.  We discuss edges here, but an identical description carries over
to sites.
Assign each edge on a graph --- make it the lattice $Z^d$ for concreteness ---
a random
variable chosen independently from the others and from a common continuous
distribution, say for
specificity the uniform distribution on $[0,1]$.  We order the edges by the
values of their
associated random variables; an edge $\{x,y\}$ will be said to be of lower
order than an
edge $\{x',y'\}$ if the random variable assigned to $\{x,y\}$ is less than that
assigned to
$\{x',y'\}$.  (Note that the distribution of the random ordering does not
depend on the specific choice of a distribution for the edge variables.)

The invasion procedure can now be described as follows.  Starting from some
arbitrary initial
site $x_0$, choose the edge of lowest order connected to it.  Consider now both
sites
connected by that edge, and examine all other edges connected to them.  Again,
choose from among
those the edge of lowest order.  One now has a cluster of three sites; one
examines all
(previously unchosen) edges connected to them, and again chooses the edge of
lowest order.
Repeating this procedure {\it ad infinitum\/}, one generates an infinite
cluster, called the
{\it invasion region\/} of $x_0$.  This cluster has several interesting
properties; among
others, it exhibits the property of ``self-organized criticality'',\refto{Bak}
in that the invasion region of any site asymptotically approaches the {\it
incipient infinite
cluster\/} of the associated {\it independent\/} bond percolation
problem.\refto{perc}  That is,
the dimensionality of the invasion region far from $x_0$ approaches the fractal
dimensionality of the incipient cluster at $p_c$ in the independent bond
percolation problem
on the identical lattice.  There are other interesting, and, for our purposes,
important
properties of invasion percolation, which will be introduced as they become
relevant to our
discussion.

\vfill\eject

\figurecaptions

1.  Standard picture of a ``rugged potential''.  The vertical axis represents
energy or free
energy, depending on the context, and the horizontal axis represents an
``abstract
configurational coordinate'' $\Phi$.  At fixed temperature and timescale
denoted by $A$, the
system can explore the region of configuration space below the corresponding
horizontal solid
line.  At the same temperature but longer timescale (or the same timescale but
higher
temperature), the system can explore the region below the horizontal line $B$
(which includes the
previous region $A$).  At a still longer timescale (or higher temperature) the
system can explore
the region below the line $C$, which includes all of $A$ and $B$.  After
Palmer, \Ref{Palmer2},
Fig.~1, and Palmer and Stein, \Ref{PS}, Fig.~3.

2. A sketch of the so-called ``incipient infinite cluster'', which is not an
infinite cluster at
all but rather a sequence of infinitely many disjoint finite clusters.  In the
figure, the largest
cluster seen in window $W_i$ is (most of) $C_i$ for $i=1$ or 2.  $C_1$ and
$C_2$ are both finite.

3.  A rough sketch of the ``ponds and outlets'' picture illustrating the diode
effect.  The first
pond contains the starting site $x_0$.  Arrows indicate the large-scale
direction of motion; once the
process leaves a given pond, it does not return.  The values of the $b_n$
decrease as $n$ increases;
$b_n$ controls the height of the minimal barrier confining the system to pond
$n$, as described in
the text.

\endfigurecaptions

\vfill\eject

\references
\mark{References}
\refis{Bak} P.~Bak, C.~Tang, and K.~Wiesenfeld, \prl 59, 381, 1987.

\refis{BY} K. Binder and A.P. Young, \rmp 58, 801, 1986.

\refis{Campbell} I.A.~Campbell, \journal J. Phys. (Paris) Lett., 46, L1159,
1985.

\refis{Chamberlin} R.V.~Chamberlin, M.~Hardiman, L.A.~Turkevich, and R.~Orbach,
\prb 25,
6720, 1982.

\refis{Chayes} J.T.~Chayes, L.~Chayes, and R.~Durrett, \jpa 20, 1521, 1987.

\refis{C2N} J.T.~Chayes, L.~Chayes, and C.M.~Newman, unpublished.

\refis{DL} B.~Derrida and J.M.~Luck, \prb 28, 7183, 1983; D.S.~Fisher, \pra 30,
960, 1984; and
J.~Bricmont and A.~Kupiainen, \prl 66, 1689, 1991 and \journal Comm. Math.
Phys., 142, 345, 1991.

\refis{erg} Of course, in a simple model like RWRE, when diffusion takes
over, there is, strictly speaking, no (normalizable) Gibbs measure for
the system and thus no possibility for equating time averages and thermal
averages.  Ergodicity in this context simply means that the system is
no longer contained in a restricted part of the total state space but is free
to diffuse; the
``ergodic time'' could equally well be called the ``containment time''.

\refis{FH}  In the droplet picture of Fisher and Huse (D.S.~Fisher and
D.A.~Huse,
\prb 38, 386, 1988), external application of an infinitesimal field does
completely
change the {\it thermodynamic\/} state of the spin glass.  However, this result
is not
germane to the present discussion; it holds for infinite systems and infinite
times.  The
size of the domain which is flipped with application of an applied field
diverges as the
field strength goes to zero.  It is therefore unlikely that this effect can
affect the
dynamical behavior of the system on short timescales.

\refis{Grimmett} G.~Grimmett, {\sl Percolation\/}, (Springer-Verlag, New York,
1989).

\refis{Guy} C.N.~Guy, \journal J.~Phys.~F, 7, 1505, 1977.

\refis{Hammersley} J.M.~Hammersley, in {\it Methods in Computational
Physics\/}, vol. I,
eds.~B. Adler, S.~Fernbach, and M.~Rotenberg, (Academic Press, New York, 1963),
pp.~281--298.

\refis{hypercube} That is, the relation derived in I between the RWRE and
invasion percolation
(which is defined in Appendix A of this paper) is valid for any graph ${\cal
G}$. However, the
{\it structure\/} of invasion percolation, and hence of the RWRE, could depend
significantly on
the nature of ${\cal G}$.  If the underlying dynamics is diffusion on a rugged
landscape in
$R^d$, then (as discussed in I) the appropriate graph is one whose vertices
correspond to the
locally stable states in $R^d$.  Although this graph will not be exactly $Z^d$,
we expect the
same qualitative structure for its invasion percolation.  If the underlying
dynamics is that of a
disordered Ising model with state space $\{-1,1\}^N$, then the nature of the
appropriate ${\cal
G}$ is a nontrivial problem in its own right, which we do not address in this
paper.

\refis{inv} R.~Lenormand and S.~Bories, \journal C.R. Acad. Sci., 291, 279,
1980; R.~Chandler,
J.~Koplick, K.~Lerman, and J.F.~Willemsen, \journal J. Fluid Mech., 119, 249,
1982;
D.~Wilkinson and J.F.~Willemsen, \jpa 16, 3365, 1983.

\refis{Jackle} J.J\"ackle,\journal Phil. Mag. B, 44, 533, 1981.

\refis{Kesten} H.~Kesten, \journal Prob. Theory Rel. Fields, 73, 369, 1986.

\refis{KK} R.W.~Knitter and J.S.~Kouvel, \journal J. Magn. Magn. Mat., 21,
L316, 1980.

\refis{Krey} U.~Krey, \journal J. Magn. Magn. Mat., 6, 27, 1977.

\refis{models}  For a review of some of this work as applied to spin glasses,
see \Ref{BY}.
For an extensive discussion of ``rugged landscapes'' and their applications,
see
S.A.~Kauffman, {\sl Origins of Order\/}, (Oxford University Press, Oxford,
1993).

\refis{Narayan}  Another viewpoint might be that application of a field
``tilts'' the
energy surface, in which case an analysis similar to that of Narayan and Fisher
might be
applicable:  O.~Narayan and D.S.~Fisher, \prb 49, 9469, 1994.

\refis{NKH} S.~Nagata, P.H.~Keesom, and H.R.~Harrison, \prb 19, 1633, 1979.

\refis{Note}  This should not be confused with the statement in ordinary BE
that there
exists some highest barrier, $\Delta F_{\rm max}$, confining the system, and
that after
escape occurs over $\Delta F_{\rm max}$, dynamical processes are ergodic.  In
that case,
the barriers are assumed to grow logarithmically with time until the system
escapes over
$\Delta F_{\rm max}$.  In model B, our assertion applies to the broken ergodic
regime
itself; after initial transients, barriers are roughly constant, with small
fluctuations.  Furthermore, there is no clear analogue in our models to
$\Delta F_{\rm
max}$.

\refis{NS1} C.M.~Newman and D.L.~Stein, ``Random Walk in a Strongly
Inhomogeneous
Environment and Invasion Percolation'', {\it Annales de L\'{}Institut Henri
Poincar\'e\/},
to appear.

\refis{NS2} C.M.~Newman and D.L.~Stein, \prl 72, 2286, 1994.

\refis{Palmer1}  R.G.~Palmer, \journal Adv. Phys., 31, 669, 1982.

\refis{Palmer2} R.G.~Palmer, in {\sl Heidelberg Colloquium on Spin Glasses},
eds.~J.L.~van
Hemmen and I.~Morgenstern (Springer-Verlag, Berlin, 1983), pp.~234--251.

\refis{perc} D. Stauffer and A. Aharony, {\sl Introduction to Percolation
Theory; 2nd Edition}
(London:  Taylor and Francis), 1992.

\refis{PS} R.G.~Palmer and D.L.~Stein, in {\sl Relaxations in Complex
Systems\/},
eds.~K.L.~Ngai and G.B.~Wright (U.S. GPO, Washington 1985), pp.~253--259.

\refis{PV} G.~Papanicolaou and S.R.S.~Varadhan, in {\it Statistics and
Probability: Essays in
Honor of C.R. Rao\/}, eds.~G.~Kallianpur, P.R.~Krishnaiah, and J.K.~Ghosh,
(North-Holland,
Amsterdam, 1982), pp.~547-552; C.~Kipnis and S.R.S.~Varadhan, \journal Comm.
Math. Phys., 104, 1,
1986; and A.~De Masi, P.A.~Ferrari, S.~Goldstein, and D.W.~Wick, \jsp 55, 787,
1989.

\refis{PWA1} P.W.~Anderson, in {\it Amorphous Magnetism II\/}, eds.~R.A.~Levy
and
R.~Hasegawa, (Plenum, NY, 1977), pp.~1--16.

\refis{PWA2} P.W.~Anderson, in {\it Ill Condensed Matter\/}, eds.~R.~Balian,
R.~Maynard, and
G.~Toulouse (North Holland, Amsterdam, 1979), pp.~159--261.

\refis{recur}  In the context of the invasion process, the second statement is
valid in the sense
that the invasion region starting from anywhere in the second pond will not
contain any sites
from the first pond.  In the context of the strongly inhomogeneous RWRE,
versions of both
statements are valid as a consequence of the results of I.  Here, for example,
is a formulation
which is valid even for $d=2$ and does not conflict with recurrence:  for any
fixed $N$ and with
probability approaching one as $\beta$ tends to $\infty$, the RWRE will visit
all sites in the
second through $N$th ponds before it either returns to the first pond or
crosses a bond of larger
value than the first outlet.

\refis{Refrigier} P.~Refrigier, E.~Vincent, J.~Hamman, and M.~Ocio, \journal J.
Phys.
(Paris), 48, 1533, 1987.

\refis{RWRE} F.~Solomon, \journal Ann. Prob., 3, 1, 1975; and H.~Kesten,
M.V.~Kozlov,
and F.~Spitzer, \journal Comp. Math., 30, 145, 1975.

\refis{Sibani} P.~Sibani and K.-H.~Hoffmann, \prl 63, 2853, 1989;
K.-H.~Hoffmann and
P.~Sibani, \journal Z. Phys. B, 80, 429, 1990.

\refis{Sinai} Y.G.~Sinai, \journal Theory Prob. Appl., 27, 256, 1982; and
M.~Bramson and
R.~Durrett, \journal Comm. Math. Phys., 119, 199, 1988.

\refis{Stein87} D.L.~Stein, in {\sl Chance and Matter}, ed.\ J. Souletie \etal\
(North-Holland,
Amsterdam, 1987), pp.~577 -- 610.

\refis{Stein88} R.G.~Palmer, in {\sl Lectures in the Sciences of Complexity\/},
ed.~D.L.~Stein (Addison-Wesley, Reading, MA, 1989), pp.~275--300.

\refis{TT} J.-L. Tholence and R.~Tournier, \journal J. Phys. (Paris), 35,
C4-229, 1974.

\refis{vEvH} A.C.D.~van Enter and J.L.~van Hemmen, \pra 29, 355, 1984.
\endreferences
\endpage

\end